\documentclass{aa}  
\usepackage{graphicx}
\usepackage{txfonts}
\usepackage{longtable}
\usepackage{natbib} 
\begin{document}
   \title{Rotational spectrum of deuterated and $^{15}$N ethyl cyanides: CH$_3$CHDCN and CH$_2$DCH$_2$CN and of CH$_3$CH$_2$C$^{15}$N.
   \thanks{Tables \ref{pln15}, \ref{plchdcn}, \ref{pldinplane} and \ref{pldoutofplane} are only available in electronic form at the CDS via anonymous ftp to cdsarc.u-strasbg.fr (130.79.125.5) or via http://cdsweb.u-strasbg.fr/cgi-bin/qcat?J/A+A/}
}
  \author{L. Margul\`es \inst{1}
         \and
          R. Motiyenko \inst{1} %\thanks{\emph{Present address: Institute of Radio Astronomy of NASU, Chervonopraporna str, 4, 61002 Kharkov, Ukraina}} 
         \and
         K. Demyk \inst{2}
          \and
         B. Tercero \inst{3}
         \and
         J. Cernicharo \inst{3}
          \and
	  M. Sheng \inst{4}
	  \and
	   M. Weidmann \inst{4}
	   \and
	   J. Gripp \inst{4}
         \and
          H. M\"{a}der \inst{4}
 	  \and 
          J. Demaison \inst{1}
          }

  \offprints{K. Demyk, \email{karine.demyk@cesr.fr}}

   \institute{Laboratoire de Physique des Lasers, Atomes et Mol\'ecules, UMR CNRS 8523,
              Universit\'e Lille 1, F-59655 Villeneuve d'Ascq Cedex, France
       \and
     Centre d'Etude Spatiale des Rayonnements, UMR CNRS 5187, Universit\'e de Toulouse (UPS), 
                    9 avenue du Colonel Roche, BP 4346, F-31028 Toulouse, France
      \and  
      Departamento de Astrof\'isica Molecular e Infrarroja, Instituto de Estructura de la Materia, CSIC, Calle Serrano 121, E-28006 Madrid, Spain
             \and
             Universit\"{a}t Kiel, Institut f\"{u}r Physikalische Chemie
             Olshausenstrasse 40, D-24098 Kiel , Germany
 }
   \date{Received 2 September 2008; accepted 13 October 2008}
% \abstract{}{}{}{}{} 
% 5 {} token are mandatory
   \abstract
  % {blabla} leave it empty if necessary  
   {Ethyl cyanide is an abundant molecule in hot molecular clouds. Its rotational spectrum is very dense and several hundreds of rotational transitions within the ground state have been identified in molecular clouds in the 40 - 900 GHz frequency range. Lines from $^{13}$C isotopically substituted ethyl cyanide were identified in Orion. }
   {To enable the search and the possible detection of other isotopologues of ethyl cyanide in interstellar objects, we have studied the rotational spectrum of deuterated ethyl cyanide: CH$_2$DCH$_2$CN (in-plane and out-of-plane) and CH$_3$CHDCN and the spectrum of $^{15}$N substituted ethyl cyanide CH$_3$CH$_2$C$^{15}$N.  Using these experimental data, we have searched for these species in Orion.}
   {The rotational spectrum of each species in the ground state was measured in the microwave and millimeter-submillimeter wavelength range using a waveguide Fourier transform spectrometer (8 - 17 GHz) and a source-modulated spectrometer employing backward-wave oscillators (BWOs) (150 - 260 and 580 - 660 GHz). More than 300 lines were identified for each species, for J values in the range 71-80 and K$\mathrm{_a}$ values in the range 28-31 depending on the isotopologues. The experimental spectra were analyzed using a Watson's Hamiltonian in the A-reduction.}
   {From the fitting procedure, accurate spectroscopic constants were derived for each of the species. These new sets of spectroscopic constants enable us to predict reliably the rotational spectrum (lines frequencies and  intensities) in the 4-1000 GHz frequency range and for J and K$\mathrm{_a}$ up to 80 and 31, respectively. Combined with IRAM 30 m antenna observations of Orion, this experimental study allowed us to detect $^{15}$N substituted ethyl cyanide CH$_3$CH$_2$C$^{15}$N for the first time in Orion. The derived column density and rotational temperature are 10$^{13}$ cm$^{-2}$ and 150 K  for the plateau and 3 $\times$10$^{14}$ cm$^{-2}$ and 300 K for the hot core. The deuterated species were searched for but were not detected. The upper limit to the column density of each deuterated isotopologues was 10$^{14}$ cm$^{-2}$.   }
   {}


  \titlerunning{Rotational spectrum of deuterated and $^{15}$N ethyl cyanide}
   \authorrunning{Margul\`es et al.}
   \maketitle
%
%________________________________________________________________

\section{Introduction} 

Ethyl cyanide, CH$_3$CH$_2$CN, is an asymmetric top molecule with a large dipole moment ($\mathrm{\mu_a}$ = 3.85 D and $\mathrm{\mu_b}$ = 1.23 D) that exhibits a dense and intense rotational spectrum. It is present in the densest parts of hot molecular cores, where it is proposed to form on dust grains. Several hundreds of lines of CH$_3$CH$_2$CN in the ground state have been observed towards hot cores such as Orion, Sgr B2 and W51 \citep{miao1997,liu2001} but also toward low mass star-forming regions \citep{cazaux2003,remijan2006}. It has a high abundance of the order of 10$^{15}$ - 10$^{17}$ cm$^{-2}$ depending on the sources \citep{miao1997,remijan2004,remijan2006}. Transitions from vibrationally excited ethyl cyanide have also been observed in Sgr B2 \citep{merhinger2004} and in W51 e2 \citep{demyk2008}. Numerous lines from $^{13}$C-substituted ethyl cyanide have been  detected in Orion Irc2 \citep{demyk2007}. 

All of these observations show that the unidentified lines observed in spectral surveys of molecular clouds are partly due to transitions from known species in vibrationally excited states or from isotopologues of known species. The most promising carriers of these transitions are the so-called {\it interstellar  weeds}, i.e. molecules, such as ethyl cyanide, which have a dense and intense rotational spectrum and/or low-frequency vibrational modes, such as methyl formate HCOOCH$_3$, dimethyl-ether CH$_3$OCH$_3$, or methanol CH$_3$OH. With the increase in sensitivity and frequency coverage that will be achieved with instruments such as HIFI onboard the Herschel Space Observatory and ALMA, the identification of these {\it U-lines} will become crucial to the search for new molecules but also to obtain important information about the physical and chemical conditions prevailing in the observed sources. 

An enormous amount of experimental work must be undertaken to complete the actual knowledge of the rotational spectra of low-frequency vibrational mode and of the isotopologues of abundant interstellar molecules. When they exist, measurements are indeed often limited to low frequencies and therefore cannot be used to predict reliable line frequencies and intensities in the Herschel and ALMA frequency ranges.

In this context, following up our study on $^{13}$C-substituted ethyl cyanide \citep{demyk2007}, we present an experimental study of the rotational spectrum of deuterated ethyl cyanide, CH$_2$DCH$_2$CN in-plane, CH$_2$DCH$_2$CN out-of-plane, CH$_3$CHDCN, and $^{15}$N substituted ethyl cyanide, CH$_3$CH$_2$C$^{15}$N.
These species were studied 30 years ago by \citet{mader1973} and \citet{heise1976}. For each species, about 30 low-J (J $\leq$ 6) rotational transitions were recorded and assigned in the 4 - 40 GHz frequency range. The rotational constants and quadrupole coupling constants derived from these studies were used to extend the measurements to higher frequency.
The isotopologues synthesis and the experimental setup are described in Sect. \ref{exp}. The analysis of the measured rotational transitions, the resulting spectroscopic parameters, and the prediction of rotational spectrum for each isotopologue in the 8 - 1000 GHz range are presented in Sect. \ref{analysis}. The detection of CH$_3$CH$_2$C$^{15}$N in Orion and the search for the deuterated species is presented in Sect. \ref{astro}.

\section{Experimental setup}
\label{exp}

The deuterated ethyl cyanides, CH$_2$DCH$_2$CN and CH$_3$CHDCN, have been prepared from the corresponding ethyl iodides CH$_2$DCH$_2$I and CH$_3$CHDI (98 atom \%, C/D/N isotopes, Pointe Claire, Canada) and normal potassium cyanide. KC$^{15}$N (98 atom\%, Aldrich, Taufkirchen, Germany) and normal ethyl iodide were used for the preparation of the $^{15}$N- isotopologue.
The potassium cyanide was dispersed in a solution of ethyl iodide in triethylene glycol and the mixture was stirred and heated slowly to 110\ $^{\circ}$C. Potassium cyanide and ethyl iodide were used in a molar ratio 1.25 : 1.00, and the concentration of the ethyl iodide solution was 4 mol/l. Nitrogen was bubbled through the mixture and the ethyl cyanide was isolated in an ice-cooled trap.
The yield of this modified Kolbe reaction \citep{organikum,mader1973} was 75\%. The product was distilled and controlled spectroscopically. The only detectable impurity was a small amount of the corresponding isonitrile. \\

In Kiel, the measurements in the centimeter-wave range were performed by means of waveguide Fourier-transform microwave-spectroscopy \citep{sarka1997}. A spectrometer in the range of about 8-17 GHz was used, employing an oversized X-band sample cell with a rectangular waveguide of quadratic cross-section and 12 m length \citep{kruger1993}. The experiments were carried out at ambient temperature and at gas pressures of ca. 0.1 Pa. Experimental transition frequencies were obtained from an analysis of the frequency-domain signals; these were derived following Fourier transformation of the transient emission signal, using a peak finder routine that determined line-center frequencies to an accuracy typically superior to 20 kHz, depending on the strength of the line. In the case of observed line splittings due to methyl internal rotation (AE-splittings) and/or the $^{14}$N-nuclear quadrupole coupling, the experimental peak frequencies were corrected to obtain hypothetical unsplit line frequencies (without nuclear quadrupole hyperfine structure).\\

The millimeter spectra were recorded in Lille in the spectral range 150 - 250 and 580 - 660 GHz. The sources were Russian Istok backward-wave oscillators (BWO). They were phase locked on an harmonic from a HP synthesizer. Up to 250 GHz, the signal from the synthesizer was directly mixed onto a Russian planar Schottky diode with part of the signal from the BWO. From 500 GHz to 650 GHz, an active sextupler from millitech (75 - 100 GHz) and a Schottky planar diode placed in a parabolic structure (from Virginia Diodes Inc.) optimized for this range were used. The detector was an InSb liquid He-cooled bolometer from QMC. To improve the sensitivity of the spectrometer, the sources were frequency-modulated at 5 kHz. The absorption cell was a stainless steel tube (6 cm diameter, 110 cm long), and the pressure that we used during measurements was 2.6 Pa (26 $\mu$bar). The accuracy of isolated lines was superior to 30 kHz.

\section{Spectral analysis and line predictions}
\label{analysis}

Ethyl cyanide and its isotopologues are prolate asymmetric top molecules. The main isotopologue, CH$_3$CH$_2$CN, has a large dipole moment ($\mathrm{\mu_a}$ = 3.83 D, $\mathrm{\mu_b}$ = 1.23 D; \citet{heise1976}), which was used to calculate the lines intensities of the four studied isotopologues, since it was found that the rotation of the principal axes system upon isotopic substitution does not induce significant variation in the dipole moment. The experimental  spectrum of each isotopologue is very dense and intense. It contains lines from the main isotopologue, which may be present as a trace in the samples. It also exhibits lines from the first low-frequency vibrationally excited states (the CH$_3$ torsion mode and the CCN bending mode). Consequently, some lines in the measured spectra are blended or distorded and are therefore not used in the analysis. $^{14}$N-nuclear quadrupole coupling and the internal rotation of the CH$_3$ group introduce splitting of the lines. However, these effects were not taken into account in the analysis, since these splittings are not observed in the millimeter range, and in the microwave region the unsplit line frequencies are used when splitting is observed. \\

The spectral analysis is a step-by-step process with permanent interaction between measurements and theory. First of all, we used the spectroscopic parameters derived from previous experimental studies at low frequency (4 - 40 GHz) to provide a first prediction of the line positions at low frequency and for low J and K$\mathrm{_a}$ value for each species. We used the experimental work from \citet{heise1976} and \citet{mader1976} for CH$_3$CH$_2$C$^{15}$N and CH$_2$DCH$_2$CN, respectively. The quartic distortional constants missing ($\delta_\mathrm{J}$ and $\delta_\mathrm{K}$) were fixed to the values of the normal species. In the absence of any experimental data, the rotational constants of CH$_3$CHDCN, were calculated using ab initio calculation at the level B3LYP/6-31G* with Gaussian 03 \citep{gaussian}. Based on these predictions, new lines were identified in new measurements performed in the 8 - 17 GHz range. All of the measured lines were then fitted to derive a new and more accurate set of spectroscopic constants. These constants were then used to derive a new prediction at higher frequency and for a higher value of J and K$\mathrm{_a}$. New lines were measured and added to the fit step by step until the fit of the measured lines and the precision of the derived spectroscopic parameters were sufficiently good for a reliable prediction to be made for the desired frequency range and value of quantum numbers. For the fitting procedure of the measured lines and for the predictions, we have used a Watson's Hamiltonian using A-reduction in I$\mathrm{^r}$ representation \citep{watson77}. The S reduction was also attempted but without significant improvement (as for the $^{13}$C species, \citet{demyk2007}). The molecular parameters were determined by fitting the experimental frequencies using the iteratively re-weighted least squares fitting method \citep{hamilton1992,bakri2002}. The objective of this method was to derive suitable weights using the residuals of a previous iteration. It had the advantage of being more robust than the standard least squares fitting methods and automatically rejecting most of the misassigned lines.\\

The number of measured lines, the maximum value of J and K$\mathrm{_a}$, and the standard deviation of the fits are presented in Table \ref{meas} for the four studied species: CH$_3$CH$_2$C$^{15}$N, CH$_2$DCH$_2$CN in-plane, CH$_2$DCH$_2$CN out-of-plane, and CH$_3$CHDCN. For the CH$_3$CH$_2$C$^{15}$N and CH$_3$CHDCN species, it is possible to reduce the standard deviation of the fit slightly by using higher order centrifugal distortion constants; but they are only marginally determined and they worsen the precision of the predictions. They were therefore no longer considered. The spectroscopic parameters derived from the fitting procedure are presented in Table \ref{paramA}. The list of measured lines are accessible in the online section as Tables \ref{mln15}, \ref{mlchd}, \ref{mldinplane}, and \ref{mldoutofplane} for CH$_3$CH$_2$C$^{15}$N, CH$_3$CHDCN, CH$_2$DCH$_2$CN in-plane, and CH$_2$DCH$_2$CN out-of-plane, respectively. For each measured line, the tables indicate its assignment (quantum numbers), the measured frequency, the difference between the observed and calculated frequency, the line strength, the dipole component, and the energy of the lower level. \\

The frequency range of the measurements (8 - 660 GHz) and the value of the quantum numbers of the identified lines (J $\le$ 80 and K$\mathrm{_a} \le$ 31) are suitable for astronomical studies. Ethyl cyanide is present in hot cores and has a rather high temperature, in the 100 - 300 K range, as its isotopologues should also have. For such temperatures (100 and 300 K), the most intense rotational transitions of ethyl cyanide occur around 238 and 407 GHz, respectively, corresponding to J values of 25 and 45, respectively. The set of spectroscopic parameters derived for each species thus allows us to predict reliably the line frequencies and the band intensity of the transitions in the spectral range suitable for interstellar detection. For each species we have calculated a prediction of the rotational spectrum in the 8 - 1000 GHz range for J $\le$ 100 and K$\mathrm{_a}$ $\le$ 35 . A short sample of the predictions is shown in the online section for each isotopologues (Tables \ref{pln15}, \ref{plchdcn}, \ref{pldinplane} and \ref{pldoutofplane}), the entire tables are available in electronic form at the CDS. The tables indicate the quantum numbers of the transition, the calculated frequency and uncertainty, the line strength, the dipole component and the energy of the lower level. The calculated error (third column in the tables) is estimated from the accuracy of the spectroscopic parameters derived by the fitting procedure. However, to get a more realistic estimation of the error, it must be multiplied by a factor 3 for the strongest lines to 10, for the weakest lines. 
The precision on the line frequency is good enough for line identification in interstellar spectra. For the lines that are the most suitable for detection, i.e., the strongest lines having J value up to $\sim$ 50-60 and a low K$_a$ value, the error on the predicted frequencies is a few hundred kHz. The error is larger for the weakest lines and increases as J and K$_a$ become larger, i.e., as the frequency increases.\\

\begin{table*}
\caption{\label{meas}Spectroscopic measurements of the ground vibrational state of CH$_3$CHDCN, CH$_2$DCH$_2$CN and CH$_3$CH$_2$C$^{15}$N}
\centering
\begin{tabular}{ l c c c c c}
\hline
\hline
		 \multicolumn{3}{r}{Number of measured lines} & J max & K$\mathrm{_a}$ max & standard\\
							    &	8-39 GHz	& 150-660 GHz &  & & deviation$^a$ (kHz) \\
\hline
CH$_2$DCH$_2$CN in-plane  		& 45	& 415 & 80 & 28 & 27 \\	
CH$_2$DCH$_2$CN out-of plane	& 58 & 496 & 74 & 29 & 28 \\
CH$_3$CHDCN					& 55 & 222 & 71 & 29 & 57$^b$ \\
CH$_3$CH$_2$C$^{15}$N		& 67 & 270 & 78 & 31 & 64$^b$ \\
\hline
\hline
\end{tabular}\\
$^a$ calculated from the median of absolute deviations\\
 $^b$ For the millimeter wave transitions, the standard deviation of the microwave transitions is 16 kHz.
\end{table*}

\begin{table*}
\caption{\label{paramA}Spectroscopic constants of the ground vibrational state of CH$_3$CHDCN, CH$_2$DCH$_2$CN and CH$_3$CH$_2$C$^{15}$N}
\centering
\begin{tabular}{ l r r r r}
\hline
\hline
\multicolumn{1}{c}{} & \multicolumn{1}{c}{CH$_2$DCH$_2$CN in-plane}  &  \multicolumn{1}{c}{CH$_2$DCH$_2$CN out-of-plane} & \multicolumn{1}{c}{CH$_3$CHDCN}  &  \multicolumn{1}{c}{CH$_3$CH$_2$C$^{15}$N}\\
\hline
A /MHz 	     	     	        	&   27651.28641(85)	& 25022.90806(62) &  24449.0602(15) & 27541.8853(13) 	\\
B /MHz 	      	     	           	&   4425.144021(82) & 4583.488572(88) &  4661.39671(21) & 4574.82824(19) 	\\
C /MHz 	      	                     	&   4000.798104(76) & 4110.239468(83) &  4155.45941(20) & 4119.44781(19) 	\\
$\Delta_\mathrm{J}$ /kHz	   &	       2.565394(33) &      3.122672(31) &      2.839982(82) &     2.893555(79) \\
$\Delta_\mathrm{_{JK}}$ /kHz  &       -45.12388(92)	&     -39.91080(59) &      -34.0809(14) &      -46.1359(13) 	\\
$\Delta_\mathrm{_K}$ /kHz	   &        551.5425(70) &       405.8505(38) &       365.720(24) &       547.265(14) \\
$\delta_\mathrm{_j}$ /Hz  	   &          550.855(23) &        703.046(20) &        656.296(27) &       635.249(38) \\
\hline
\hline
\end{tabular}\\
Standard deviation of the fits are given Table \ref{meas}, see text for more details. Uncertainties given in parenthesis are in units of the last digit given and 1 times the standard deviation. 
\end{table*}

\onllongtab{3}{\begin{longtable}{ r r r r r r r r r r r }
\caption{\label{mln15}Measured transitions of the ground 
vibrational state of CH$_3$CH$_2$C$^{15}$N}\\
\hline
\hline
\multicolumn{6}{c}{Transition} & \multicolumn{1}{c}{Obs. Freq.} & \multicolumn{1}{c}{obs.- calc.} & \multicolumn{1}{c}{S}& \multicolumn{1}{c}{Dipole} & \multicolumn{1}{c}{E$\mathrm{_l}$} \\
\multicolumn{1}{c}{J'} & \multicolumn{1}{c}{K'$\mathrm{_a}$} & \multicolumn{1}{c}{K'$\mathrm{_c}$} & \multicolumn{1}{c}{J"} & \multicolumn{1}{c}{K"$\mathrm{_a}$} & \multicolumn{1}{c}{K"$\mathrm{_c}$} & \multicolumn{1}{c}{(MHz)} & \multicolumn{1}{c}{(MHz)} & & & \multicolumn{1}{c}{(cm$^{-1}$)} \\
 & & & & & & & & & & \\
\endfirsthead
\multicolumn{11}{c}%
{ \tablename\ \thetable{} Measured transitions of CH$_3$CH$_2$C$^{15}$N-- continued from previous page} \\
\hline
\hline
\multicolumn{6}{c}{Transition} & \multicolumn{1}{c}{Obs. Freq.} & \multicolumn{1}{c}{obs.- calc.} & \multicolumn{1}{c}{S}& \multicolumn{1}{c}{Dipole} & \multicolumn{1}{c}{E$\mathrm{_l}$} \\
\multicolumn{1}{c}{J'} & \multicolumn{1}{c}{K'$\mathrm{_a}$} & \multicolumn{1}{c}{K'$\mathrm{_c}$} & \multicolumn{1}{c}{J"} & \multicolumn{1}{c}{K"$\mathrm{_a}$} & \multicolumn{1}{c}{K"$\mathrm{_c}$} & \multicolumn{1}{c}{(MHz)} & \multicolumn{1}{c}{(MHz)} & & & \multicolumn{1}{c}{(cm$^{-1}$)} \\
\hline 
\endhead
\hline 
\multicolumn{11}{r}{{\it{continued on next page}}} \\ 
\endfoot
\endlastfoot
\hline
13	&	2	&	11	&	13	&	2	&	12	&	8439.616	&	0.003	&	5.5270	&	a	&	29.44	\\
27	&	6	&	22	&	28	&	5	&	23	&	8579.609	&	-0.004	&	4.3669	&	b	&	137.32	\\
1	&	0	&	1	&	0	&	0	&	0	&	8694.274	&	0.010	&	1.0000	&	a	&	0.00		\\
27	&	6	&	21	&	28	&	5	&	24	&	8864.577	&	0.008	&	4.3657	&	b	&	137.31	\\
31	&	4	&	27	&	31	&	4	&	28	&	9029.287	&	-0.011	&	0.8976	&	a	&	156.65	\\
30	&	5	&	26	&	29	&	6	&	23	&	9223.293	&	-0.017	&	4.7882	&	b	&	154.18	\\
61	&	7	&	54	&	61	&	7	&	55	&	9230.917	&	-0.004	&	1.3312	&	a	&	587.92	\\
7	&	2	&	5	&	8	&	1	&	8	&	9270.407	&	0.020	&	1.0280	&	b	&	10.93	\\
\hline
\hline
\end{longtable}
}

\onllongtab{4}{\begin{longtable}{ r r r r r r r r r r r }
\caption{\label{mlchd}Measured transitions of the ground 
vibrational state of CH$_3$CHDCN}\\
\hline
\hline
\multicolumn{6}{c}{Transition} & \multicolumn{1}{c}{Obs. Freq.} & \multicolumn{1}{c}{obs.- calc.} & \multicolumn{1}{c}{S}& \multicolumn{1}{c}{Dipole} & \multicolumn{1}{c}{E$\mathrm{_l}$} \\
\multicolumn{1}{c}{J'} & \multicolumn{1}{c}{K'$\mathrm{_a}$} & \multicolumn{1}{c}{K'$\mathrm{_c}$} & \multicolumn{1}{c}{J"} & \multicolumn{1}{c}{K"$\mathrm{_a}$} & \multicolumn{1}{c}{K"$\mathrm{_c}$} & \multicolumn{1}{c}{(MHz)} & \multicolumn{1}{c}{(MHz)} & & & \multicolumn{1}{c}{(cm$^{-1}$)} \\
 & & & & & & & & & & \\
\endfirsthead
\multicolumn{11}{c}%
{ \tablename\ \thetable{} Measured transitions of CH$_3$CHDCN-- continued from previous page} \\
\hline
\hline
\multicolumn{6}{c}{Transition} & \multicolumn{1}{c}{Obs. Freq.} & \multicolumn{1}{c}{obs.- calc.} & \multicolumn{1}{c}{S}& \multicolumn{1}{c}{Dipole} & \multicolumn{1}{c}{E$\mathrm{_l}$} \\
\multicolumn{1}{c}{J'} & \multicolumn{1}{c}{K'$\mathrm{_a}$} & \multicolumn{1}{c}{K'$\mathrm{_c}$} & \multicolumn{1}{c}{J"} & \multicolumn{1}{c}{K"$\mathrm{_a}$} & \multicolumn{1}{c}{K"$\mathrm{_c}$} & \multicolumn{1}{c}{(MHz)} & \multicolumn{1}{c}{(MHz)} & & & \multicolumn{1}{c}{(cm$^{-1}$)} \\
 & & & & & & & & & & \\
\hline 
\endhead
\hline 
\multicolumn{11}{r}{{\it{continued on next page}}} \\ 
\endfoot
\endlastfoot
\hline
21	&	4	&	17	&	20	&	5	&	16	&	8145.638	&	-0.015	&	3.1333	&	b	&	78.57	\\
36	&	5	&	31	&	36	&	5	&	32	&	8324.713	&	-0.032	&	1.1806	&	a	&	213.36	\\
63	&	8	&	55	&	63	&	8	&	56	&	8399.871	&	-0.011	&	1.6500	&	a	&	637.87	\\
7	&	1	&	6	&	6	&	2	&	5	&	8464.923	&	0.012	&	1.2346	&	b	&	8.85	\\
9	&	3	&	7	&	10	&	2	&	8	&	8618.098	&	-0.003	&	1.3927	&	b	&	18.98	\\
12	&	2	&	10	&	12	&	2	&	11	&	8717.220	&	0.009	&	0.5901	&	a	&	25.56	\\
54	&	7	&	47	&	54	&	7	&	48	&	8805.418	&	-0.005	&	1.4898	&	a	&	471.24	\\
45	&	6	&	39	&	45	&	6	&	40	&	8810.297	&	-0.001	&	1.3333	&	a	&	329.72	\\
\hline
\hline
\end{longtable}
}

\onllongtab{5}{\begin{longtable}{ r r r r r r r r r r r }
\caption{\label{mldinplane}Measured transitions of the ground 
vibrational state of CH$_2$DCH$_2$CN in-plane}\\
\hline
\hline
\multicolumn{6}{c}{Transition} & \multicolumn{1}{c}{Obs. Freq.} & \multicolumn{1}{c}{obs.- calc.} & \multicolumn{1}{c}{S}& \multicolumn{1}{c}{Dipole} & \multicolumn{1}{c}{E$\mathrm{_l}$} \\
\multicolumn{1}{c}{J'} & \multicolumn{1}{c}{K'$\mathrm{_a}$} & \multicolumn{1}{c}{K'$\mathrm{_c}$} & \multicolumn{1}{c}{J"} & \multicolumn{1}{c}{K"$\mathrm{_a}$} & \multicolumn{1}{c}{K"$\mathrm{_c}$} & \multicolumn{1}{c}{(MHz)} & \multicolumn{1}{c}{(MHz)} & & & \multicolumn{1}{c}{(cm$^{-1}$)} \\
 & & & & & & & & & & \\
\endfirsthead
\multicolumn{11}{c}%
{ \tablename\ \thetable{} Measured transitions of CH$_2$DCH$_2$CN in-plane- continued from previous page} \\
\hline
\hline
\multicolumn{6}{c}{Transition} & \multicolumn{1}{c}{Obs. Freq.} & \multicolumn{1}{c}{obs.- calc.} & \multicolumn{1}{c}{S}& \multicolumn{1}{c}{Dipole} & \multicolumn{1}{c}{E$\mathrm{_l}$} \\
\multicolumn{1}{c}{J'} & \multicolumn{1}{c}{K'$\mathrm{_a}$} & \multicolumn{1}{c}{K'$\mathrm{_c}$} & \multicolumn{1}{c}{J"} & \multicolumn{1}{c}{K"$\mathrm{_a}$} & \multicolumn{1}{c}{K"$\mathrm{_c}$} & \multicolumn{1}{c}{(MHz)} & \multicolumn{1}{c}{(MHz)} & & & \multicolumn{1}{c}{(cm$^{-1}$)} \\
 & & & & & & & & & & \\
\hline 
\endhead
\hline 
\multicolumn{11}{r}{{\it{continued on next page}}} \\ 
\endfoot
\endlastfoot
\hline
1	&	0	&	1	&	0	&	0	&	0	&	8425.939	&	0.007	&	1.0000	&	a	&	0.00	\\
32	&	4	&	28	&	32	&	4	&	29	&	8545.382	&	0.014	&	0.8744	&	a	&	161.32	\\
6	&	1	&	5	&	6	&	1	&	6	&	8904.345	&	0.033	&	0.3101	&	a	&	6.53	\\
17	&	4	&	14	&	18	&	3	&	15	&	9080.706	&	0.031	&	2.8265	&	b	&	55.26	\\
14	&	2	&	12	&	14	&	2	&	13	&	9580.530	&	0.060	&	0.5108	&	a	&	32.58	\\
12	&	1	&	12	&	11	&	2	&	9	&	9611.583	&	-0.026	&	1.2987	&	b	&	21.79	\\
23	&	3	&	20	&	23	&	3	&	21	&	9736.492	&	-0.005	&	0.6882	&	a	&	84.74	\\
12	&	3	&	9	&	13	&	2	&	12	&	9782.272	&	0.007	&	1.9814	&	b	&	28.66	\\
\hline
\hline
\end{longtable}
}

\onllongtab{6}{\begin{longtable}{ r r r r r r r r r r r }
\caption{\label{mldoutofplane}Measured transitions of the ground  vibrational state of CH$_2$DCH$_2$CN out-of-plane}\\
\hline\hline
\multicolumn{6}{c}{Transition} & \multicolumn{1}{c}{Obs. Freq.} & \multicolumn{1}{c}{obs.- calc.} & \multicolumn{1}{c}{S}& \multicolumn{1}{c}{Dipole} & \multicolumn{1}{c}{E$\mathrm{_l}$} \\
\multicolumn{1}{c}{J'} & \multicolumn{1}{c}{K'$\mathrm{_a}$} & \multicolumn{1}{c}{K'$\mathrm{_c}$} & \multicolumn{1}{c}{J"} & \multicolumn{1}{c}{K"$\mathrm{_a}$} & \multicolumn{1}{c}{K"$\mathrm{_c}$} & \multicolumn{1}{c}{(MHz)} & \multicolumn{1}{c}{(MHz)} & & & \multicolumn{1}{c}{(cm$^{-1}$)} \\
 & & & & & & & & & & \\
\endfirsthead
\multicolumn{11}{c}%
{ \tablename\ \thetable{} Measured transitions of CH$_2$DCH$_2$CN out-of-plane- continued from previous page} \\
\hline
\hline
\multicolumn{6}{c}{Transition} & \multicolumn{1}{c}{Obs. Freq.} & \multicolumn{1}{c}{obs.- calc.} & \multicolumn{1}{c}{S}& \multicolumn{1}{c}{Dipole} & \multicolumn{1}{c}{E$\mathrm{_l}$} \\
\multicolumn{1}{c}{J'} & \multicolumn{1}{c}{K'$\mathrm{_a}$} & \multicolumn{1}{c}{K'$\mathrm{_c}$} & \multicolumn{1}{c}{J"} & \multicolumn{1}{c}{K"$\mathrm{_a}$} & \multicolumn{1}{c}{K"$\mathrm{_c}$} & \multicolumn{1}{c}{(MHz)} & \multicolumn{1}{c}{(MHz)} & & & \multicolumn{1}{c}{(cm$^{-1}$)} \\
 & & & & & & & & & & \\
\hline 
\endhead
\hline 
\multicolumn{11}{r}{{\it{continued on next page}}} \\ 
\endfoot
\endlastfoot
\hline
22	&	4	&	18	&	21	&	5	&	17	&	8505.160	&	-0.007	&	3.3575	&	b	&	84.32	\\
6	&	2	&	4	&	7	&	1	&	7	&	8520.324	&	0.006	&	0.8664	&	b	&	8.58	\\
1	&	0	&	1	&	0	&	0	&	0	&	8693.720	&	0.004	&	1.0000	&	a	&	0.00	\\
29	&	4	&	25	&	29	&	4	&	26	&	8713.615	&	0.001	&	0.9523	&	a	&	137.58	\\
10	&	1	&	10	&	9	&	2	&	7	&	8777.717	&	-0.014	&	1.1077	&	b	&	15.88	\\
38	&	5	&	33	&	38	&	5	&	34	&	8862.059	&	0.000	&	1.1166	&	a	&	232.89	\\
38	&	9	&	30	&	39	&	8	&	31	&	9150.310	&	0.088	&	5.8015	&	b	&	270.62	\\
10	&	3	&	7	&	11	&	2	&	10	&	9258.390	&	-0.006	&	1.5568	&	b	&	21.87	\\
\hline
\hline
\end{longtable}
}

\onltab{7}{\begin{table*}
\caption{\label{pln15}Sample table of the predicted transitions of the ground  vibrational state of CH$_3$CH$_2$C$^{15}$N}
\centering
\begin{tabular}{ c c c c c c }
\hline 
\hline
Transition    						    & Calc. Frequency & Calc. Error & S & Dipole & E$\mathrm{_l}$  \\
J' (K'$\mathrm{_a}$, K'$\mathrm{_c}$) - J" (K"$\mathrm{_a}$, K"$\mathrm{_c}$) & (MHz)               & (MHz)        &    &           & (cm$^{-1}$) \\
\hline
3 (1, 3) - 2 (1, 2)	&	25395.710 	&	0.05	&	2.6666	&	A 	&	1.62	\\
4 (1, 3) - 4 (0, 4)	&	25554.136 	&	0.05	&	4.3003	&	B 	&	2.90	\\
3 (0,	3) - 2 (0, 2)	&	26055.720 	&	0.05	&	2.9994	&	A 	&	0.87	\\
3 (2,	2) - 2 (2, 1)	&	26083.623 	&	0.05	&	1.6667	&	A 	&	3.96	\\
3 (2,	1) - 2 (2, 0)	&	26110.418 	&	0.05	&	1.6667	&	A 	&	3.96	\\
3 (1,	2) - 2 (1, 1)	&	26761.502 	&	0.05	&	2.6666	&	A 	&	1.67	\\
\end{tabular}
\end{table*}
}

\onltab{8}{\begin{table*}
\caption{\label{plchdcn}Sample table of the predicted transitions of the ground  vibrational state of CH$_3$CHDCN}
\centering
\begin{tabular}{ c c c c c c }
\hline 
\hline
Transition    						    & Calc. Frequency & Calc. Error & S & Dipole & E$\mathrm{_l}$  \\
J' (K'$\mathrm{_a}$, K'$\mathrm{_c}$) - J" (K"$\mathrm{_a}$, K"$\mathrm{_c}$) & (MHz)               & (MHz)        &    &           & (cm$^{-1}$) \\
\hline
3 (1,	3) - 2 (1, 2	)	&	25685.775	&	0.04	&	2.6666	&	A 	&	1.53	\\
6 (1,	5) - 6	 (0, 6	)	&	25869.041	&	0.04	&	5.6814	&	B 	&	6.15	\\
3 (0,	3) - 2	 (0, 2	)	&	26411.989	&	0.04	&	2.999	&	A 	&	0.88	\\
3 (2,	2) - 2	 (2, 1	)	&	26451.080	&	0.04	&	1.6667	&	A 	&	3.56	\\
3 (2,	1) - 2	 (2, 0	)	&	26489.353	&	0.04	&	1.6667	&	A 	&	3.56	\\
3 (1,	2) - 2	 (1, 1	)	&	27203.193	&	0.04	&	2.6666	&	A 	&	1.58	\\
\end{tabular}
\end{table*}
}

\onllongtab{9}{\begin{table}
\caption{\label{pldinplane}Sample table of the predicted transitions of the ground  vibrational state of CH$_2$DCH$_2$CN in-plane}
\centering
\begin{tabular}{ c c c c c c }
\hline 
\hline
Transition    						    & Calc. Frequency & Calc. Error & S & Dipole & E$\mathrm{_l}$  \\
J' (K'$\mathrm{_a}$, K'$\mathrm{_c}$) - J" (K"$\mathrm{_a}$, K"$\mathrm{_c}$) & (MHz)               & (MHz)        &    &           & (cm$^{-1}$) \\
\hline
3 (0,	3) - 2	 (0, 2	) &	25254.521	&	0.03	&	2.9995	&	A 	&	0.84	\\
3 (2,	2) - 2	 (2, 1	) &	25278.632	&	0.03	&	1.6667	&	A 	&	3.97	\\
3 (2,	1) - 2	 (2, 0	) &	25301.661	&	0.03	&	1.6667	&	A 	&	3.97	\\
4 (1,	3) - 4	 (0, 4	) &	25630.983	&	0.03	&	4.3159	&	B 	&	2.81	\\
3 (1,	2) - 2	 (1, 1	) &	25910.592	&	0.03	&	2.6666	&	A 	&	1.65	\\
5 (1,	4) - 5	 (0, 5	) &	26781.045	&	0.03	&	5.1502	&	B 	&	4.21	\\
\end{tabular}
\end{table}
}

\onllongtab{10}{\begin{table}
\caption{\label{pldoutofplane}Sample table of the predicted transitions of the ground  vibrational state of CH$_2$DCH$_2$CN out-of-plane}
\centering
\begin{tabular}{ c c c c c c }
\hline 
\hline
Transition    						    & Calc. Frequency & Calc. Error & S & Dipole & E$\mathrm{_l}$  \\
J' (K'$\mathrm{_a}$, K'$\mathrm{_c}$) - J" (K"$\mathrm{_a}$, K"$\mathrm{_c}$) & (MHz)               & (MHz)        &    &           & (cm$^{-1}$) \\
\hline
3 (1,	3) - 2 (1, 2	)	&	25366.325	&	0.03	&	2.6666	&	A 	&	1.54	\\
3 (0,	3) - 2 (0, 2	)	&	26048.385	&	0.03	&	2.9992	&	A 	&	0.87	\\
3 (2,	2) - 2 (2, 1	)	&	26081.805	&	0.03	&	1.6667	&	A 	&	3.63	\\
6 (1,	5) - 6 (0, 6	)	&	26083.982	&	0.03	&	5.7568	&	B 	&	6.07	\\
3 (2,	1) - 2 (2, 0	)	&	26114.266	&	0.03	&	1.6667	&	A 	&	3.63	\\
3 (1,	2) - 2 (1, 1	)	&	26785.701	&	0.03	&	2.6666	&	A 	&	1.58	\\
\end{tabular}
\end{table}
}

\section{CH$_3$CH$_2$C$^{15}$N detection in Orion}
\label{astro}

\subsection{Astronomical observation}

The observations were carried out using the IRAM 30 m radio telescope during 2004 September (3 mm and 1.3 mm), 2005 March (2 mm), 2005 April (3 mm and 1.3 mm). We acquired data for the entire spectral range detectable by the 30-m recievers. The four SiS receivers operating at 3, 2 and 1.3 mm were used simultaneously. Each receiver was tuned to a single sideband with image rejections within 20-27 dB (3 mm receivers), 12-16 dB (2 mm receivers), and 13 dB (1.3 mm receivers).

System temperatures were  100-350 K for the 3 mm receivers, 200-500 K for the 2 mm  receivers, and 200-800 K for the 1.3 mm receivers, depending on the particular frequency, weather conditions, and source elevation. The intensity scale was calibrated using two absorbers at different temperatures and using the Atmospheric Transmission Model  \citep{cer85, par01}. 

Pointing and focus were regularly monitored by observing the nearby quasars 0420-014 and 0528+134. The observations were completed in the balanced wobbler-switching mode with a wobbling frequency of 0.5 Hz and a beam throw in azimuth of $\pm$240''.
The backends used were two filter banks with 512$\times$1 MHz channels and a correlator providing two 512 MHz bandwidths and 1.25 MHz resolution. We performed a spectral-line survey, for which the central frequencies were chosen in a systematic way: from 80 GHz to 115.5 GHz for the 3 mm domain; from 130.25 GHz to 176.75 GHz for 2 mm; from 197 to 141 GHz for 1.3 mm (low frequency) and from 141.25 to 281.75 GHz for the 1.3 mm domain (high frequency), in steps of 500 MHz. We pointed toward the (survey) position $\alpha$ = 5$^h$ 35$^m$ 14.5$^s$, $\delta$ = -5$^{\circ}$ 22' 30.0'' (J2000.0), corresponding to IRc2. The detailed procedure used for the analysis of the line survey is described in Tercero et al. (in preparation).

\subsection{Astronomical modeling}

In agreement with previous observations of Orion, four clearly defined  kinematic regions with quite different physical and chemical conditions  \citep{bla87, bla96} are implied by the observed LSR  velocities and line widths:  (i) the narrow ($\lesssim$5 km s$^{-1}$ line width)   feature at v$_{\mathrm{LSR}}$ $\simeq$ 9 km s$^{-1}$, forming a N-S \textit{extended ridge} or ambient cloud; (ii) a compact and quiescent  region, \textit{compact ridge}, (v$_{\mathrm{LSR}}$ $\simeq$ 8 km s$^{-1}$,  $\Delta$v $\simeq$ 3 km s$^{-1}$), identified for the first time by  \citet{joh84}; (iii) the more turbulent and compact \textit{plateau} (v$_{\mathrm{LSR}}$ $\simeq$ 6-10 km s$^{-1}$, $\Delta$v $\gtrsim$25 km s$^{-1}$); (iv) the \textit{hot core} component (v$_{\mathrm{LSR}}$ $\simeq$  3-5 km s$^{-1}$, $\Delta$v $\lesssim$10-15 kms$^{-1}$), first observed  in ammonia emission \citep{mor80}.\\

In modeling the emission from the $^{15}$N isotopologue of ethyl cyanide, we found that, as for the $^{13}$C isotopologues \citep{demyk2007}, a sum of two components is sufficient to reproduce all line intensities and profiles reasonably well: the hot core component and the plateau. We assumed LTE for both components. For the hot core, a column density of 3$\times$10$^{14}$ cm$^{-2}$, a line width of 5 km s$^{-1}$, and a rotational temperature of 300 K were the most suitable parameters for reproducing the bulk of the CH$_3$CH$_2$C$^{15}$N emission. A broad velocity component is was also required to reproduce the observations accurately, corresponding to the plateau for which we obtain a column density of 1$\times$10$^{13}$ cm$^{-2}$, a line width of 20 km s$^{-1}$, and a rotational temperature of 150 K. For the hot core component, we assumed a source diameter of 7'' with uniform brightness temperature and optical depth over its extent at 3'' from the pointed position (the observation were pointed towards IRc2, while the CN bearing species appears to originate in a small region 3'' North). For the plateau component, we assumed a size for the source of 30''.

\begin{figure*}
\centering
\includegraphics[scale=1]{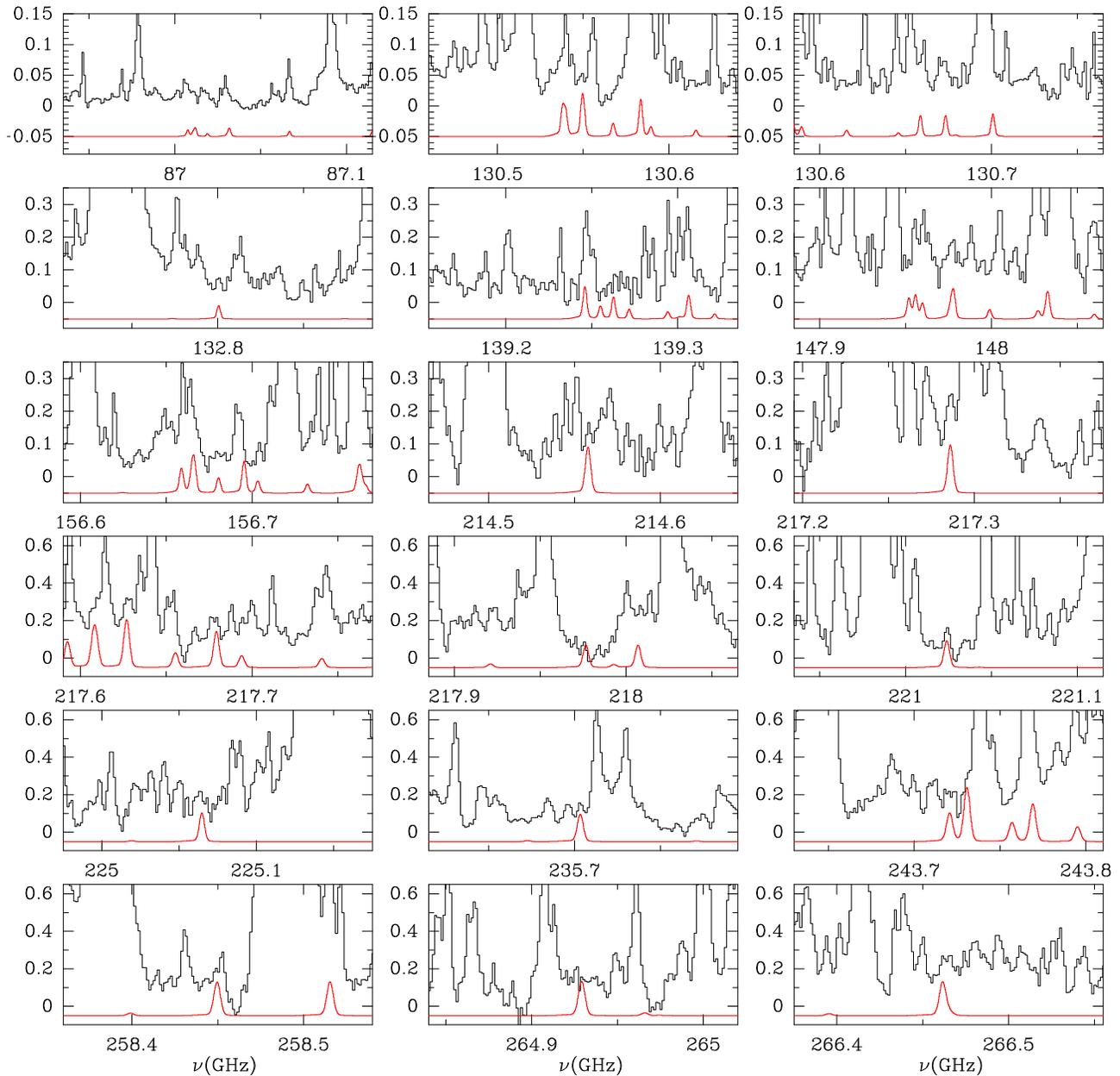}
\caption{\label{orion} $^{15}$N-ethyl cyanide isotopologues detection in Orion. The spectra are in units of main beam temperature (T$_{mb}$). The histogram spectra offset with respect to each other are the observations (black curve) and the model (smooth red line).
}
\label{15n}
\end{figure*}

The lines of CH$_3$CH$_2$C$^{15}$N are weak and many of them are heavily blended with lines from other species. However, no missing lines were found in the coverage of our line survey of Orion and the lines of CH$_3$CH$_2$C$^{15}$N free of blending appear at the correct frequencies and with the correct intensities. Although our modeling lacks a robust analysis, the difference between model and observed intensities is always below 20\%. Our modeling of the different components is straightforward, although we emphasize that we modeled the main isotope, for which strong lines are observed, with the same model and the lines are well reproduced. Hence, we are confident about the results for the isotope $^{15}$N, in particular that the frequency of all detected lines differ from laboratory measurements by less than 0.5 MHz.  Hence, our assignment of the lines shown in Fig. \ref{15n} and in Table \ref{15n-obs} to the $^{15}$N isotopologue is fairly secure. The procedure that we used to identify the carriers of the weak lines in Orion, many of which remain unassigned at present, is the most suitable for this source since it permits to confirm whether there are no missing lines of the molecules for which we are looking.

Fig. \ref{15n} shows selected observed lines of the $^{15}$N isotopologue with model results. Table \ref{15n-obs} indicates the model predictions, observed peak intensities and frequencies, and predicted frequencies from the rotational constants obtained in this paper, for all lines of CH$_3$CH$_2$C$^{15}$N that were not strongly blended with other lines. The differences between the intensity of the model and the peak intensity of the observed lines were mostly due to the contribution from many other molecular species (the strong overlap with other lines ensures that it is difficult to provide a good baseline for the weak lines of CH$_3$CH$_2$C$^{15}$N).

Using the column density derived for the $^{13}$C isotopologues (1.6$\times$10$^{15}$ and 6$\times$10$^{13}$ cm$^{-2}$ for the hot core and the plateau, respectively; see \citealt{demyk2007}), we derive an isotopic ratio $^{13}$C/$^{15}$N of between 5-6, in agreement with the solar isotopic abundance ($^{13}$C/$^{15}$N (solar) $\simeq$ 6) and strengthening our identification of $^{15}$N-ethyl cyanide in Orion. The detailed modeling of ethyl cyanide including the main isotopologue, the vibrationally excited states and the detected isotopologues, will be published elsewhere (Tercero et al. in preparation).\\

Deuterated ethyl cyanide has not been detected in this line survey above the line confusion limit. Assuming the same physical conditions as those derived for CH$_3$CH$_2$C$^{15}$N, we have derived an upper limit to the column densities of CH$_2$DCH$_2$CN (in-plane and out-of-plane) of 1$\times$10$^{14}$ cm$^{-2}$ in both species.

\onllongtab{11}{\begin{longtable}{  c r r r r r r} 
\caption{\label{15n-obs}Observed transitions of CH$_3$CH$_2$C$^{15}$N (without high blend) in the frequency range of the Orion KL survey. Column 1 indicates the transition, column 2 provides the assumed rest frequency, column 3 the line strength, column 4 the energy of the upper level, column 5 the observed (centroid) frequencies assuming that the radial velocities relative to LSR are 5 km s$^{-1}$, column 6 the observed peak line temperature of main beam, and column 7 indicates the main beam temperature derived from the model.
 }\\
\hline\hline
Transitions & Pred. Freq. & S$\mathrm{_{ij}}$ & E$\mathrm{_u}$/k & Obs. Freq. & Obs.  T$\mathrm{_{mb}}$ & Model T$\mathrm{_{mb}}$\\
J$\mathrm{_{K_a,K_c}}$ & (MHz) & & (K) & (MHz) & (K) &  (K) \\
\endfirsthead
\multicolumn{7}{c}%
{ \tablename\ \thetable{} Observed transitions of CH$_3$CH$_2C^{15}$N-- continued from previous page} \\
\hline
\hline
Transitions & Pred. Freq. & S$\mathrm{_{ij}}$ & E$\mathrm{_u}$/k & Obs. Freq. & Obs.  T$\mathrm{_{mb}}$ & Model T$\mathrm{_{mb}}$\\
J$\mathrm{_{K_a,K_c}}$ & (MHz) & & (K) & (MHz) & (K) &  (K) \\
\hline 
\endhead
\hline 
\multicolumn{7}{r}{{\it{continued on next page}}} \\ 
\endfoot
\endlastfoot
\hline
10$_{6,5}$-9$_{6,4}$ & 87007.366 & 6.40 & 63.0 & 87007.2 & 0.044 & 0.011\\
10$_{6,4}$-9$_{6,3}$ & 87007.366 & 6.40 & 63.0 & \footnotemark[1] & & \\
10$_{7,4}$-9$_{7,3}$ & 87010.616 & 5.10 & 77.4 & 87010.3 & 0.024 & 0.015\\
10$_{7,3}$-9$_{7,2}$ & 87010.616 & 5.10 & 77.4 & \footnotemark[1] & & \\
10$_{5,6}$-9$_{5,5}$ & 87011.913 & 7.50 & 50.8 & \footnotemark[1] & & \\
10$_{5,5}$-9$_{5,4}$ & 87011.919 & 7.50 & 50.8 & \footnotemark[1] & & \\
10$_{8,2}$-9$_{8,1}$ & 87018.779 & 3.60 & 94.1 & 87019.2 & 0.026 & 0.005\\
10$_{8,3}$-9$_{8,2}$ & 87018.779 & 3.60 & 94.1 & \footnotemark[1] & & \\
10$_{3,7}$-9$_{3,6}$ & 87114.933 & 9.10 & 33.0 & 87114.2 & 0.070\footnotemark[2] & 0.009\\
10$_{2,8}$-9$_{2,7}$ & 87815.417 & 9.60 & 27.5 & 87814.4 & 0.015 & 0.010\\
10$_{1,9}$-9$_{1,8}$ & 88914.019 & 9.89 & 24.6 & 88914.3 & 0.015 & 0.011\\
11$_{1,11}$-10$_{1,10}$ & 92801.263 & 10.9 & 27.9 & 92803.2\footnotemark[2] & 0.020 & 0.013\\
11$_{0,11}$-10$_{0,10}$ & 94252.099 & 11.0 & 27.3 & 94251.3 & 0.040\footnotemark[3] & 0.014\\
11$_{2,10}$-10$_{2,9}$ & 95397.849 & 10.6 & 32.0 & 95398.2 & 0.010 & 0.013\\
11$_{4,7}$-10$_{4,6}$ & 95750.117 & 9.55 & 45.4 & 95750.2 & 0.009 & 0.015\\
11$_{3,8}$-10$_{3,7}$ & 95868.920 & 10.2 & 37.6 & 95868.2 & 0.015 & 0.012\\
11$_{2,9}$-10$_{2,8}$ & 96765.809 & 10.6 & 32.2 & 96764.2 & 0.024 & 0.014\\
12$_{0,12}$-11$_{0,11}$ & 102578.600 & 12.0 & 32.2 & 102577.1 & 0.042 & 0.018\\
12$_{4,9}$-11$_{4,8}$ & 104469.699 & 10.7 & 50.4 & 104470.1 & 0.059 & 0.017\\
12$_{4,8}$-11$_{4,7}$ & 104472.697 & 10.7 & 50.4 & \footnotemark[1] & & \\
12$_{3,10}$-11$_{3,9}$ & 104517.172 & 11.2 & 42.6 & 104517.2 & 0.027 & 0.017\\
12$_{2,10}$-11$_{2,9}$ & 105746.402 & 11.7 & 37.3 & 105748.1 & 0.036 & 0.019\\
12$_{1,11}$-11$_{1,10}$ & 106509.810 & 11.9 & 34.4 & 106511.1 & 0.041\footnotemark[2] & 0.020\\
13$_{0,13}$-12$_{0,12}$ & 110865.655 & 13.0 & 37.6 & 110867.0 & 0.032\footnotemark[2] & 0.024\\
13$_{2,12}$-12$_{2,11}$ & 112621.740 & 12.7 & 42.4 & 112621.0 & 0.027 & 0.024\\
13$_{6,8}$-12$_{6,7}$ & 113127.839 & 10.2 & 78.0 & 113128.0 & 0.130 & 0.034\\
13$_{6,7}$-12$_{6,6}$ & 113127.839 & 10.2 & 78.0 & \footnotemark[1] & & \\
13$_{8,6}$-12$_{8,5}$ & 113129.990 & 8.08 & 109.1 & \footnotemark[1] & & \\
13$_{8,5}$-12$_{8,4}$ & 113129.990 & 8.08 & 109.1 & \footnotemark[1] & & \\
13$_{4,9}$-12$_{4,8}$ & 113200.301 & 11.8 & 55.8 & 113201.0 & 0.107 & 0.021\\
13$_{3,11}$-12$_{3,10}$ & 113245.049 & 12.3 & 48.0 & 113243.9 & 0.095\footnotemark[2] & 0.023\\
%2 mm
15$_{7,9}$-14$_{7,8}$ & 130537.903 & 11.7 & 104.6 & 130539.7 & 0.149\footnotemark[2] & 0.054\\
15$_{7,8}$-14$_{7,7}$ & 130537.903 & 11.7 & 104.6 & \footnotemark[1] & & \\
15$_{8,8}$-14$_{8,7}$ & 130539.798 & 10.7 & 121.3 & \footnotemark[1] & & \\
15$_{8,7}$-14$_{8,6}$ & 130539.798 & 10.7 & 121.3 & \footnotemark[1] & & \\
15$_{6,10}$-14$_{6,9}$ & 130549.183 & 12.6 & 90.2 & 130549.8 & 0.052 & 0.070\\
15$_{6,9}$-14$_{6,8}$ & 130549.187 & 12.6 & 90.2 & \footnotemark[1] & & \\
15$_{9,6}$-14$_{9,5}$ & 130550.372 & 9.60 & 140.1 & \footnotemark[1] & & \\
15$_{9,7}$-14$_{9,4}$ & 130550.372 & 9.60 & 140.1 & \footnotemark[1] & & \\
15$_{10,6}$-14$_{10,5}$ & 130567.307 & 8.33 & 161.2 & 130567.7 & 0.049 & 0.022\\
15$_{10,5}$-14$_{10,4}$ & 130567.307 & 8.33 & 161.2 & \footnotemark[1] & & \\
15$_{12,3}$-14$_{12,2}$ & 130615.586 & 5.40 & 209.9 & 130614.7 & 0.042 & 0.010\\
15$_{12,4}$-14$_{12,3}$ & 130615.586 & 5.40 & 209.9 & \footnotemark[1] & & \\
15$_{4,11}$-14$_{4,10}$ & 130673.185 & 13.9 & 67.9 & 130672.7 & 0.073 & 0.034\\
15$_{1,14}$-14$_{1,13}$ & 132670.810 & 14.9 & 52.3 & 132670.7 & 0.130\footnotemark[4] & 0.042\\
15$_{2,13}$-14$_{2,12}$ & 132800.381 & 14.7 & 55.1 & 132799.7 & 0.075 & 0.041\\
16$_{8,9}$-15$_{8,8}$ & 139245.508 & 12.0 & 127.9 & 139247.0 & 0.280\footnotemark[2] & 0.098\\
16$_{8,8}$-15$_{8,7}$ & 139245.508 & 12.0 & 127.9 & \footnotemark[1] & & \\
16$_{7,10}$-15$_{7,9}$ & 139246.220 & 12.9 & 111.3 & \footnotemark[1] & & \\
16$_{7,9}$-15$_{7,8}$ & 139246.220 & 12.9 & 111.3 & \footnotemark[1] & & \\
16$_{6,11}$-15$_{6,10}$ & 139262.522 & 13.8 & 96.8 & 139263.3 & 0.086 & 0.067\\
16$_{6,10}$-15$_{6,9}$ & 139262.529 & 13.8 & 96.8 & \footnotemark[1] & & \\
16$_{10,7}$-15$_{10,6}$ & 139271.671 & 9.75 & 167.9 & 139270.8 & 0.078 & 0.029\\
16$_{10,6}$-15$_{10,5}$ & 139271.671 & 9.75 & 167.9 & \footnotemark[1] & & \\
16$_{5,12}$-15$_{5,11}$ & 139306.059 & 14.4 & 84.6 & 139305.8 & 0.293\footnotemark[2] & 0.073\\
16$_{5,11}$-15$_{5,10}$ & 139306.594 & 14.4 & 84.6 & \footnotemark[1] & & \\
16$_{12,4}$-15$_{12,3}$ & 139321.451 & 7.00 & 216.6 & 139321.9 & 0.045 & 0.015\\
16$_{12,5}$-15$_{12,4}$ & 139321.451 & 7.00 & 216.6 & \footnotemark[1] & & \\
16$_{1,15}$-15$_{1,14}$ & 141313.454 & 15.9 & 59.1 & 141314.5 & 0.060 & 0.051\\
17$_{0,17}$-16$_{0,16}$ & 143750.890 & 16.9 & 62.8 & 143750.6 & 0.18\footnotemark[5]1 & 0.056\\
17$_{8,10}$-16$_{8,9}$ & 147951.800 & 13.2 & 135.0 & 147955.6 & 0.256 & 0.064\\
17$_{8,9}$-16$_{8,8}$ & 147951.800 & 13.2 & 135.0 & \footnotemark[1] & & \\
17$_{7,11}$-16$_{7,10}$ & 147955.646 & 14.1 & 118.4 & \footnotemark[1] & & 0.074\\
17$_{7,10}$-16$_{7,9}$ & 147955.646 & 14.1 & 118.4 & \footnotemark[1] & & \\
17$_{9,9}$-16$_{9,8}$ & 147959.709 & 12.2 & 153.9 & \footnotemark[1] & & 0.050\\
17$_{9,8}$-16$_{9,7}$ & 147959.709 & 12.2 & 153.9 & \footnotemark[1] & & \\
17$_{10,7}$-16$_{10,6}$ & 147976.004 & 11.1 & 175.0 & 147975.5 & 0.170 & 0.093\\
17$_{10,8}$-16$_{10,7}$ & 147976.004 & 11.1 & 175.0 & \footnotemark[1] & & \\
17$_{6,12}$-16$_{6,11}$ & 147977.798 & 14.9 & 103.9 & 147977.5 & 0.192\footnotemark[2] & \\
17$_{6,11}$-16$_{6,10}$ & 147977.813 & 14.9 & 103.9 & \footnotemark[1] & & \\
17$_{5,13}$-16$_{5,12}$ & 148032.124 & 15.5 & 91.7 & 148030.7 & 0.245 & 0.083\\
17$_{5,12}$-16$_{5,11}$ & 148033.058 & 15.5 & 91.7 & \footnotemark[1] & & \\
17$_{4,14}$-16$_{4,13}$ & 148139.616 & 16.1 & 81.7 & 148137.0 & 0.104\footnotemark[2] & 0.053\\
17$_{3,15}$-16$_{3,14}$ & 148147.371 & 16.5 & 74.0 & 148146.9 & 0.155\footnotemark[2] & 0.055\\
17$_{4,13}$-16$_{4,12}$ & 148174.997 & 16.1 & 81.7 & 148174.4 & 0.158\footnotemark[6] & 0.051\\
17$_{3,14}$-16$_{3,13}$ & 148810.665 & 16.5 & 74.1 & 148810.5 & 0.080 & 0.055\\
17$_{1,16}$-16$_{1,15}$ & 149910.805 & 16.9 & 66.3 & 149909.5 & 0.177\footnotemark[2] & 0.061\\
17$_{2,15}$-16$_{2,14}$ & 150846.477 & 16.8 & 69.1 & 150847.5 & 0.135 & 0.060\\
18$_{1,18}$-17$_{1,17}$ & 151161.544 & 17.9 & 70.3 & 151170.6 & 0.185\footnotemark[7] & 0.064\\
18$_{9,10}$-17$_{9,9}$ & 156664.727 & 13.5 & 161.4 & 156663.4 & 0.298\footnotemark[2] & 0.116\\
18$_{9,9}$-17$_{9,8}$ & 156664.727 & 13.5 & 161.4 & \footnotemark[1] & & \\
18$_{7,12}$-17$_{7,11}$ & 156666.249 & 15.3 & 125.9 & \footnotemark[1] & & \\
18$_{7,11}$-17$_{7,10}$ & 156666.250 & 15.3 & 125.9 & \footnotemark[1] & & \\
18$_{10,9}$-17$_{10,8}$ & 156680.304 & 12.4 & 182.5 & 156679.4 & 0.077 & 0.047\\
18$_{10,8}$-17$_{10,7}$ & 156680.304 & 12.4 & 182.5 & \footnotemark[1] & & \\
18$_{6,13}$-17$_{6,12}$ & 156695.130 & 16.0 & 111.5 & 156693.4 & 0.196\footnotemark[2] & 0.097\\
18$_{6,12}$-17$_{6,11}$ & 156695.157 & 16.0 & 111.5 & \footnotemark[1] & & \\
18$_{4,14}$-17$_{4,13}$ & 156939.258 & 17.1 & 89.3 & 156938.4 & 0.133 & 0.060\\
18$_{3,15}$-17$_{3,14}$ & 157730.365 & 17.5 & 81.6 & 157729.4 & 0.060 & 0.065\\
18$_{2,16}$-17$_{2,15}$ & 159848.455 & 17.8 & 76.8 & 159847.4 & 0.105 & 0.071\\
19$_{2,18}$-18$_{2,17}$ & 163915.948 & 18.8 & 83.4 & 163916.3 & 0.160\footnotemark[8] & 0.077\\
19$_{1,18}$-18$_{1,17}$ & 166956.959 & 18.9 & 81.9 & 166957.3 & 0.125 & 0.082\\
20$_{2,19}$-19$_{2,18}$ & 172402.881 & 19.8 & 91.7 & 172401.6 & 0.228 & 0.088\\
20$_{8,13}$-19$_{8,12}$ & 174074.508 & 16.8 & 158.8 & 174072.8 & 0.426\footnotemark[9] & 0.166\\
20$_{8,12}$-19$_{8,11}$ & 174074.508 & 16.8 & 158.8 & \footnotemark[1] & & \\
20$_{9,12}$-19$_{9,11}$ & 174075.518 & 16.0 & 177.7 & \footnotemark[1] & & \\
20$_{9,11}$-19$_{9,10}$ & 174075.518 & 16.0 & 177.7 & \footnotemark[1] & & \\
20$_{7,14}$-19$_{7,13}$ & 174091.254 & 17.6 & 142.2 & 174090.3 & 0.345\footnotemark[10] & 0.128\\
20$_{7,13}$-19$_{7,12}$ & 174091.255 & 17.6 & 142.2 & \footnotemark[1] & & \\
%1 mm
23$_{7,17}$-22$_{7,16}$ & 200239.252 & 20.9 & 169.8 & 200241.0 & 0.143 & 0.162\\
23$_{7,16}$-22$_{7,15}$ & 200239.261 & 20.9 & 169.8 & \footnotemark[1] & & \\
23$_{12,12}$-22$_{12,11}$ & 200251.101 & 16.7 & 275.1 & 200250.9 & 0.136 & 0.064\\
23$_{12,11}$-22$_{12,10}$ & 200251.101 & 16.7 & 275.1 & \footnotemark[1] & & \\
23$_{1,22}$-22$_{1,21}$ & 200425.284 & 22.9 & 117.9 & 200423.4\footnotemark[2] & 0.152 & 0.125\\
23$_{16,7}$-22$_{16,6}$ & 200442.132 & 11.9 & 398.7 & 200441.0 & 0.016 & 0.020\\
23$_{16,8}$-22$_{16,7}$ & 200442.132 & 11.9 & 398.7 & \footnotemark[1] & & \\
23$_{4,20}$-22$_{4,19}$ & 200668.972 & 22.3 & 133.2 & 200670.9 & 0.173\footnotemark[10] & 0.110\\
23$_{2,21}$-22$_{2,20}$ & 204419.525 & 22.8 & 121.6 & 204422.9\footnotemark[2] & 0.132 & 0.128\\
25$_{0,25}$-24$_{0,24}$ & 209271.572 & 24.9 & 132.1 & 209272.1 & 0.367\footnotemark[10] & 0.138\\
24$_{2,22}$-23$_{2,21}$ & 213217.376 & 23.8 & 131.8 & 213215.8 & 0.320\footnotemark[11] & 0.139\\
25$_{2,24}$-24$_{2,23}$ & 214557.679 & 24.8 & 139.1 & 214558.3 & 0.165 & 0.140\\
26$_{1,26}$-25$_{1,25}$ & 217285.943 & 25.9 & 142.6 & 217285.7 & 0.253 & 0.147\\
26$_{0,26}$-25$_{0,25}$ & 217474.027 & 25.9 & 142.6 & 217473.3 & 0.207 & 0.147\\
25$_{9,17}$-24$_{9,16}$ & 217607.414 & 21.8 & 225.8 & 217609.5 & 0.323 & 0.228\\
25$_{9,16}$-24$_{9,15}$ & 217607.414 & 21.8 & 225.8 & \footnotemark[1] & & \\
25$_{10,16}$-24$_{10,15}$ & 217609.115 & 21.0 & 246.8 & \footnotemark[1] & & \\
25$_{10,15}$-24$_{10,14}$ & 217609.115 & 21.0 & 246.8 & \footnotemark[1] & & \\
25$_{11,15}$-24$_{11,14}$ & 217626.225 & 20.2 & 270.1 & 217626.9 & 0.320 & 0.255\\
25$_{11,14}$-24$_{11,13}$ & 217626.225 & 20.2 & 270.1 & \footnotemark[1] & & \\
25$_{8,18}$-24$_{8,17}$ & 217627.120 & 22.4 & 206.9 & \footnotemark[1] & & \\
25$_{8,17}$-24$_{8,16}$ & 217627.120 & 22.4 & 206.9 & \footnotemark[1] & & \\
25$_{7,19}$-24$_{7,18}$ & 217679.026 & 23.0 & 190.2 & 217679.4 & 0.217 & 0.191\\
25$_{7,18}$-24$_{7,17}$ & 217679.051 & 23.0 & 190.2 & \footnotemark[1] & & \\
25$_{5,21}$-24$_{5,20}$ & 217976.444 & 24.0 & 163.7 & 217975.7 & 0.086 & 0.119\\
25$_{3,22}$-24$_{3,21}$ & 221023.822 & 24.6 & 146.8 & 221024.4 & 0.162 & 0.142\\
26$_{1,25}$-25$_{1,24}$ & 225064.582 & 25.9 & 149.2 & 225063.1 & 0.218 & 0.152\\
26$_{6,21}$-25$_{6,20}$ & 226521.943 & 24.6 & 186.7 & 226524.4\footnotemark[12] & 0.244 & 0.203\\
26$_{6,20}$-25$_{6,19}$ & 226523.621 & 24.6 & 186.7 & \footnotemark[1] & & \\
26$_{5,22}$-25$_{5,21}$ & 226737.488 & 25.0 & 174.5 & 226736.8 & 0.235\footnotemark[13] & 0.127\\
26$_{4,22}$-25$_{4,21}$ & 227585.336 & 25.4 & 164.8 & 227585.5 & 0.498 & 0.139\\
28$_{1,28}$-27$_{1,27}$ & 233761.349 & 27.9 & 164.7 & 233762.4 & 0.239 & 0.163\\
28$_{0,28}$-27$_{0,27}$ & 233886.100 & 27.9 & 164.6 & 233885.3 & 0.151 & 0.164\\
27$_{3,25}$-26$_{3,24}$ & 234807.835 & 26.7 & 168.0 & 234807.4 & 0.325 & 0.154\\
27$_{10,18}$-26$_{10,17}$ & 235016.780 & 23.3 & 269.0 & 235015.4 & 0.150 & 0.146\\
27$_{10,17}$-26$_{10,16}$ & 235016.780 & 23.3 & 269.0 & \footnotemark[1] & & \\
27$_{4,24}$-26$_{4,23}$ & 235703.198 & 26.4 & 175.9 & 235703.2 & 0.156 & 0.146\\
28$_{3,26}$-27$_{3,25}$ & 243387.108 & 27.7 & 179.7 & 243385.6 & 0.207 & 0.161\\
28$_{10,19}$-27$_{10,18}$ & 243720.490 & 24.4 & 280.7 & 243720.2 & 0.217 & 0.152\\
28$_{10,18}$-27$_{10,17}$ & 243720.490 & 24.4 & 280.7 & \footnotemark[1] & & \\
28$_{9,20}$-27$_{9,19}$ & 243730.327 & 25.1 & 259.6 & 243730.2 & 0.362 & 0.289\\
28$_{9,19}$-27$_{9,18}$ & 243730.327 & 25.1 & 259.6 & \footnotemark[1] & & \\
28$_{11,18}$-27$_{11,17}$ & 243731.014 & 23.7 & 303.9 & \footnotemark[1] & & \\
28$_{11,17}$-27$_{11,16}$ & 243731.014 & 23.7 & 303.9 & \footnotemark[1] & & \\
28$_{12,17}$-27$_{12,16}$ & 243756.893 & 22.9 & 329.4 & 243756.3 & 0.200 & 0.102\\
28$_{12,16}$-27$_{12,15}$ & 243756.893 & 22.9 & 329.4 & \footnotemark[1] & & \\
28$_{6,23}$-27$_{6,22}$ & 244007.283 & 26.7 & 209.7 & 244007.2 & 0.427 & 0.149\\
28$_{6,22}$-27$_{6,21}$ & 244011.067 & 26.7 & 209.7 & 244010.2 & 0.432 & 0.148\\
29$_{6,24}$-28$_{6,23}$ & 252754.799 & 27.8 & 221.8 & 252755.2 & 0.456 & 0.139\\
29$_{6,23}$-28$_{6,22}$ & 252760.349 & 27.8 & 221.8 & 252760.1 & 0.402 & 0.137\\
29$_{5,25}$-28$_{5,24}$ & 253042.421 & 28.1 & 209.7 & 253041.7 & 0.252 & 0.147\\
29$_{4,25}$-28$_{4,24}$ & 254464.800 & 28.4 & 200.1 & 254464.2 & 0.291 & 0.160\\
31$_{1,31}$-30$_{1,30}$ & 258449.466 & 30.9 & 200.7 & 258449.2 & 0.204 & 0.180\\
30$_{6,24}$-29$_{6,23}$ & 261513.668 & 28.8 & 234.4 & 261512.9 & 0.278 & 0.140\\
30$_{2,28}$-29$_{2,27}$ & 264929.014 & 29.8 & 202.0 & 264929.9 & 0.156 & 0.182\\
32$_{0,32}$-31$_{0,31}$ & 266726.277 & 31.9 & 213.5 & 266725.0 & 0.369 & 0.183\\
31$_{3,29}$-30$_{3,28}$ & 269002.338 & 30.7 & 217.2 & 269002.9 & 0.660 & 0.175\\
31$_{8,24}$-30$_{8,23}$ & 269919.627 & 28.9 & 278.3 & 269918.9 & 0.191 & 0.224\\
31$_{8,23}$-30$_{8,22}$ & 269919.638 & 28.9 & 278.3 & \footnotemark[1] & & \\
31$_{5,27}$-30$_{5,26}$ & 270593.584 & 30.2 & 235.3 & 270592.9 & 0.262 & 0.155\\
31$_{4,28}$-30$_{4,27}$ & 270656.392 & 30.5 & 225.4 & 270655.9 & 0.390 & 0.167\\
31$_{5,26}$-30$_{5,25}$ & 270792.922 & 30.2 & 235.3 & 270791.9 & 0.357 & 0.155\\
33$_{0,33}$-32$_{0,32}$ & 274937.410 & 32.9 & 226.7 & 274936.8 & 0.277\footnotemark[14] & 0.185\\
32$_{11,22}$-31$_{11,21}$ & 278531.537 & 28.2 & 354.9 & 278530.1 & 0.736 & 0.247\\
32$_{11,21}$-31$_{11,20}$ & 278531.537 & 28.2 & 354.9 & \footnotemark[1] & & \\
32$_{10,23}$-31$_{10,22}$ & 278534.350 & 28.9 & 331.6 & \footnotemark[1] &  & \\
32$_{10,22}$-31$_{10,21}$ & 278534.350 & 28.9 & 331.6 & \footnotemark[1] & & \\
32$_{4,29}$-31$_{4,28}$ & 279370.232 & 31.5 & 238.8 & 279270.2 & 0.556 & 0.301\\
32$_{5,28}$-31$_{5,27}$ & 279371.886 & 31.2 & 248.7 & \footnotemark[1] & & \\
32$_{5,27}$-31$_{5,26}$ & 279633.253 & 31.2 & 248.7 & 279632.8 & 0.384 & 0.157\\
\hline
\hline
\footnotetext[1]{blended with the last one}
\footnotetext[2]{blended with unidentified line}
\footnotetext[3]{blended with $^{34}$SO$_2$}
\footnotetext[4]{blended with CH$_3$OCOH}
\footnotetext[5]{blended with $^{13}$CH$_2$CHCN}
\footnotetext[6]{blended with CH$_3$OCH$_3$}
\footnotetext[7]{blended with SO$^{18}$O}
\footnotetext[8]{blended with c-C$_2$H$_4$O}
\footnotetext[9]{blended with $^{13}$CH$_3$OH $\nu_t$=1 }
\footnotetext[10]{blended with CH$_3$COCH$_3$}
\footnotetext[11]{blended with H$^{15}$NCO}
\footnotetext[12]{blended with CH$_3$CH$_2$CN b type}
\footnotetext[13]{blended with CH$_3$OD}
\footnotetext[14]{blended with H$^{13}$CCCN $\nu_{7}$=2}
\end{longtable}
}

\begin{acknowledgements}
This work was supported by the Programme National "Physico Chimie du Milieu Interstellaire'' and by the European Research Training Network "Molecular Universe'' (MRTN-CT-2004-512302). R.M. has been supported by INTAS young research fellowship. The Kiel  authors thank the Land Schleswig-Holstein for financial support.
\end{acknowledgements}

\bibliographystyle{aa}
\bibliography{bibpropioDN15.bib}

\Online

\end{document}